\documentclass[prl,nofootinbib,twocolumn,superscriptaddress]{revtex4}
\def\mysection#1{{\bf #1.} }
\def\mysections#1{{\bf #1.} }
\usepackage{amssymb}
\usepackage{amsmath}
\usepackage[dvips]{graphicx}
\usepackage{longtable}
\usepackage{verbatim}
\usepackage{amsfonts}

\newcommand{\be}{\begin{equation}}
\newcommand{\ee}{\end{equation}}
\newcommand{\bea}{\begin{eqnarray}}
\newcommand{\eea}{\end{eqnarray}}
\newcommand{\beq}{\begin{equation}}
\newcommand{\eeq}{\end{equation}}
\def\beqa{\begin{eqnarray}}
\def\eeqa{\end{eqnarray}}
\newcommand{\no}{\nonumber}
\def\lsim{\mathrel{\rlap{\lower4pt\hbox{\hskip1pt$\sim$}}
    \raise1pt\hbox{$<$}}}         
\def\gsim{\mathrel{\rlap{\lower4pt\hbox{\hskip1pt$\sim$}}
    \raise1pt\hbox{$>$}}}         

\begin{document}


\vspace*{-30mm}

\title{\boldmath Combining $K^0-\overline{K^0}$ mixing 
  and $D^0-\overline{D^0}$ mixing\\ to constrain the flavor
  structure of new physics} 

\author{Kfir Blum}\email{kfir.blum@weizmann.ac.il}
\affiliation{Department of Particle Physics,
  Weizmann Institute of Science, Rehovot 76100, Israel}

\author{Yuval Grossman}\email{yg73@cornell.edu}
\affiliation{Institute for High Energy Phenomenology, Newman
  Laboratory of Elementary Particle Physics, Cornell University,
  Ithaca, NY 14853, USA}

\author{Yosef Nir}\email{yosef.nir@weizmann.ac.il}
\affiliation{Department of Particle Physics,
  Weizmann Institute of Science, Rehovot 76100, Israel}

\author{Gilad Perez}\email{gilad.perez@weizmann.ac.il}
\affiliation{Department of Particle Physics,
  Weizmann Institute of Science, Rehovot 76100, Israel}

\vspace*{1cm}

\begin{abstract}
New physics at high energy scale often contributes to
$K^0-\overline{K^0}$ and $D^0-\overline{D^0}$ mixings in an
approximately $SU(2)_L$ invariant way. In such a case, the combination
of measurements in these two systems is particularly powerful. The
resulting constraints can be expressed in terms of misalignments and
flavor splittings. 
\end{abstract}

\maketitle

\mysection{Introduction}
Measurements of flavor changing neutral current processes put strong
constraints on new physics at the TeV scale and provide a crucial
guide for model building. In particular, measurements of the mass
splitting and CP violation in the neutral $K$ system
\cite{Amsler:2008zzb}, 
\beqa\label{expk}
\Delta m_K/m_K&=&(7.01\pm0.01)\times10^{-15},\no\\
\epsilon_K&=&(2.23\pm0.01)\times10^{-3},
\eeqa
require a highly non-generic flavor structure to any such
theory. Recently, huge progress has been made in measurements of the
mass splitting and in the search for CP violation in the neutral $D$
system \cite{Barberio:2008fa}:
\beqa\label{expd}
\Delta m_D/m_D&=&(8.6\pm2.1)\times10^{-15},\no\\
A_\Gamma&=&(1.2\pm2.5)\times10^{-3}.
\eeqa
These measurements are particularly useful in constraining models
where the main flavor changing effects occur in the up sector
\cite{futuregnp}. 

By `non-generic flavor structure' we mean either alignment or
degeneracies or both. Each of the set of constraints (\ref{expk})
and (\ref{expd}) can be satisfied by aligning the new physics
contributions with specific directions in flavor space.  However,
contributions that involve only quark doublets cannot be
simultaneously aligned in both the down and the up sectors. Thus, the
combination of the measurements related to $K^0-\overline{K^0}$ mixing
(\ref{expk}) and to $D^0-\overline{D^0}$ mixing (\ref{expd}) leads to
unavoidable bounds on new physics degeneracies.

In this work, we develop the formalism that is necessary to obtain
these unavoidable bounds, explain the qualitative implications and
derive the actual quantitative constraints from the present
experimental bounds.

\mysection{Theoretical and experimental background} 
The effects of new physics at a high scale $\Lambda_{\rm NP}\gg m_W$
on low energy phenomena can be expressed in terms of an effective
Hamiltonian, composed of Standard Model (SM) fields and obeying the SM
symmetries. In particular, four-quark operators contribute to $\Delta
S=2$ and $\Delta C=2$ processes. We are interested in the operators
that involve only quark doublets:
\beq\label{nonren}
\frac{1}{\Lambda_{\rm NP}^2}
\left[z_1^K (\overline{d_L}\gamma_\mu s_L)
(\overline{d_L}\gamma^\mu s_L) +z_1^D
  (\overline{u_L}\gamma_\mu c_L)
  (\overline{u_L}\gamma^\mu c_L)\right].
\eeq
We constrain new physics by requiring that contributions of the form
(\ref{nonren}) do not exceed the experimental value of $\Delta m_K$
and the one-sigma upper bounds on $\Delta m_D$ and on CP violation in 
$D^0-\overline{D^0}$ mixing.  As concerns $\epsilon_K$, since the
SM contribution has only little uncertainties and should
be taken into account, we require that the new physics is smaller than
0.6 times the experimental bound \cite{Agashe:2005hk}. We update the
calculations of Ref. \cite{Bona:2007vi} (the details are presented in
\cite{futuregnp}) and obtain the following upper bounds on
$|z_1^K|$ and $|z_1^D|$: 
\beqa\label{expab}
|z_1^K|\leq z_{\exp}^K&=&8.8\times10^{-7}\left(\frac{\Lambda_{\rm NP}}{1\
    {\rm TeV}}\right)^2, \no\\
|z_1^D|\leq z_{\exp}^D&=&5.9\times10^{-7}\left(\frac{\Lambda_{\rm NP}}{1\
    {\rm TeV}}\right)^2,
\eeqa
and on ${\cal I}m(z_1^K)$ and ${\cal I}m(z_1^D)$:
\beqa\label{expim}
{\cal I}m(z_1^K)\leq z_{\rm exp}^{IK}&=&3.3\times10^{-9}
\left(\frac{\Lambda_{\rm NP}}{1\
    {\rm TeV}}\right)^2,\no\\
{\cal I}m(z_1^D)\leq z_{\rm exp}^{ID}&=&1.0\times10^{-7}
\left(\frac{\Lambda_{\rm NP}}{1\
    {\rm TeV}}\right)^2.
\eeqa
 
When effects of $SU(2)_L$ breaking are small, the terms that lead
to $z_1^K$ and $z_1^D$ have the form
\beq\label{xqoperator}
\frac{1}{\Lambda_{\rm NP}^2}(\overline{Q_{Li}}(X_Q)_{ij}\gamma_\mu
Q_{Lj}) (\overline{Q_{Li}}(X_Q)_{ij}\gamma^\mu Q_{Lj}),
\eeq
where $X_Q$ is an hermitian matrix. The matrix $X_Q$ provides a source
of flavor violation beyond the Yukawa matrices of the SM,
$Y_d$ and $Y_u$: 
\beq
\overline{Q_{Li}}(Y_d)_{ij}d_j\phi_d+\overline{Q_{Li}}(Y_u)_{ij}u_j\phi_u.
\eeq
Here $\phi_{d,u}$ are Higgs doublets of opposite hypercharges. (Within
the SM, $\phi_u=\tau_2\phi_d^\dagger$.) Without loss of
generality, we can choose to work in a basis where
\beq
Y_d=\lambda_d,\ \ \ Y_u=V^\dagger\lambda_u,\ \ \
X_Q=V_d^\dagger\lambda_Q V_d,
\eeq
where $\lambda_{d,u,Q}$ are diagonal real matrices, $V$ is the CKM
matrix, and $V_d$ is a unitary matrix which parameterizes the
misalignment of the operator (\ref{xqoperator}) with the down mass
basis. If we minimize the number of phases in $V$ (namely $V$ depends
on three mixing angles and a single phase), then $V_d$ depends on six
parameters: three real mixing angles and three phases.

Alternatively, we can choose to work in a basis where
\beq
Y_d=V\lambda_d,\ \ \ Y_u=\lambda_u,\ \ \
X_Q=V_u^\dagger\lambda_Q V_u.
\eeq
Here $V_u$ is a unitary matrix which parameterizes the
misalignment of the operator (\ref{xqoperator}) with the up mass
basis: $V_u = V_d V^\dagger$.

The $X_Q$-related contributions to the operators (\ref{nonren}) have
the form
\beqa\label{defzdu}
z_1^K&=&(z_{12}^d)^2\equiv[(V_d^\dagger\lambda_Q V_d)_{12}]^2,\no\\
z_1^D&=&(z_{12}^u)^2\equiv[(V_u^\dagger\lambda_Q V_u)_{12}]^2.
\eeqa

\mysection{Two generations, no CP violation}
The experimental constraints that are most relevant to our study are
related to $K^0-\overline{K^0}$ and $D^0-\overline{D^0}$ mixing, which
involve only the first two generation quarks. When studying new
physics effects, ignoring the third generation is often a good
approximation to the physics at hand. Indeed, even when the third
generation does play a role, our two generation analysis is applicable
as long as there are no strong cancellations with contributions
related to the third generation. 

In a two generation framework, $V$ depends on a single mixing angle
(the Cabibbo angle $\theta_c$), while $V_d$ depends on a single 
angle and a single phase. To understand various aspects of our
analysis, it is useful, however, to provisionally set the phase to
zero, and study only CP conserving (CPC) observables. We thus have
\beqa\label{twogen}
\lambda_Q&=&{\rm diag}(\lambda_1,\lambda_2),\no\\
V&=&\left(\begin{array}{cc}
  \cos\theta_c & \sin\theta_c \\ -\sin\theta_c & \cos\theta_c
\end{array}\right),\no\\
V_d&=&\left(\begin{array}{cc}
    \cos\theta_d & \sin\theta_d \\ -\sin\theta_d & \cos\theta_d
  \end{array}\right).
\eeqa
It is convenient to define
\beq
\lambda_{12}=\frac12(\lambda_{1}+\lambda_{2}),\ \
\delta_{12}=\frac{\lambda_{1}-\lambda_{2}}{\lambda_{1}+\lambda_{2}},\ \
\Lambda_{12}=\delta_{12}\lambda_{12}.
\eeq
Here $\lambda_{12}$ parameterizes the overall, flavor-diagonal
suppression of $X_Q$ (in particular, loop factors), while
$\delta_{12}$ parameterizes suppression that is coming from approximate
degeneracy between the eigenvalues of $X_Q$. Our analysis will yield
upper bounds on $\Lambda_{12}$, but with reasonable assumptions about
$\lambda_{12}$ they can be translated into upper bounds on the flavor
degeneracy factor $\delta_{12}$.

It is straightforward to solve for $z_1^K$ and $z_1^D$:
\beqa\label{zonecpc}
z_1^K&=&\Lambda_{12}^2\sin^22\theta_d,\no\\
z_1^D&=&\Lambda_{12}^2\sin^22(\theta_d-\theta_c).
\eeqa
We learn that the new physics contribution to $\Delta m_K$ can
be set to zero by alignment in the down sector, $\theta_d=0$,
while the contribution to $\Delta m_D$ can be set to zero by
alignment in the up sector, $\theta_d=\theta_c$. However, one cannot
set the contributions to both $\Delta m_K$ and $\Delta m_D$ to zero by
any choice of $\theta_d$. Thus, the combination of $\Delta m_K$ and
$\Delta m_D$ provides an unavoidable bound on $\Lambda_{12}$. Defining
\beq
r_{KD}\equiv\sqrt{z_{\rm exp}^K/z_{\rm exp}^D},
\eeq
the weakest bound on $\Lambda_{12}$ corresponds to
\beq
\tan2\theta_d=\frac{r_{KD}\sin2\theta_c}{1+r_{KD}\cos2\theta_c}, 
\eeq
and is given by
\beq
\Lambda_{12}^2\leq
\frac{z_{\rm exp}^{K}+z_{\rm exp}^{D}+2\sqrt{z_{\rm exp}^K z_{\rm
      exp}^D}\cos2\theta_c}{\sin^22\theta_c}.
\eeq
Using Eq. (\ref{expab}), we obtain that the weakest bound occurs at
$\sin2\theta_d\approx0.25$, and is given by
\beq\label{boucpc}
\Lambda_{12}\leq3.8\times10^{-3}
\left(\frac{\Lambda_{\rm NP}}{1\ {\rm TeV}}\right).
\eeq

We learn that for $\Lambda_{\rm NP}\leq1$ TeV, the flavor-diagonal 
and flavor-degeneracy factors should provide a suppression at least
as strong as ${\cal O}(0.004)$. To appreciate the significance of
this result, the reader should bear in mind that each of the
$\Delta m_K$ and $\Delta m_D$ bounds can be satisfied with
appropriate alignment and neither flavor-diagonal suppression nor
flavor-degeneracy. For tree level contributions with couplings of order 
one ($\lambda_{12}\sim1$), the flavor degeneracy should be stronger than
$0.004$. For loop suppression, say $\lambda_{12}\sim\alpha_2$, the
degeneracy should be stronger than $0.1$.

\mysection{Two generations, CP violation}
To incorporate CP violation (CPV), one has to employ an appropriate
formalism. We use the fact that the matrices $z^d$ and $z^u$ defined 
in Eq. (\ref{defzdu}) are hermitian. We use properties of hermitiam
matrices to parametrize $z^d$ as follows: 
\beqa
z^d&=&V_d^\dagger\lambda_Q V_d\no\\
&=&\lambda_{12}\left(\mathbb{I}+\delta_{12} V_d^\dagger\sigma_3
  V_d\right)\no\\
&\equiv&\lambda_{12}\left(\mathbb{I}+\delta_{12}\hat
  v\cdot\vec\sigma\right),
\eeqa
where $\sigma_i$ are the Pauli matrices, the $\hat v_i$ are real
with $|\hat v|=1$. The matrix $z^u$ is related to $z^d$ by a unitary
transformation involving the Cabibbo matrix. It follows that it can be
written as 
\beq
z^u=\lambda_{12}\left(\mathbb{I}+\delta_{12}\hat
  v^\prime\cdot\vec\sigma\right),
\eeq
where
\beq
\hat v^\prime=\left(\begin{array}{ccc}
    \cos2\theta_c & 0 & -\sin2\theta_c \\
    0 & 1 & 0 \\ \sin2\theta_c & 0 & \cos2\theta_c
  \end{array}\right)\hat v.
\eeq
Our formalism is motivated by the fact that it puts all CPV in $\hat
v_2$. The $\hat v_2$  parameter is the projection of $X_Q$ onto the
direction perpendicular to the $1-3$ plane where, without loss of
generality, $Y_d Y_d^\dagger$ and $Y_uY_u^\dagger$ reside. This can be
clearly seen from the expression for the Jarlskog invariant for our
framework:  
\beqa
J&=&{\rm Tr}\left\{X\left[Y_dY_d^\dag,Y_uY_u^\dag\right]\right\}\\
&=&i(y_s^2-y_d^2)(y_c^2-y_u^2)\Lambda_{12}\sin2\theta_c\ \hat
v_2. \no 
\eeqa

Using this parametrization, we obtain
\beqa
z_1^K&=&\Lambda_{12}^2(\hat v_1-i\hat v_2)^2,\\
z_1^D&=&\Lambda_{12}^2(\cos2\theta_c\hat v_1-\sin2\theta_c\hat
v_3-i\hat v_2)^2.
\eeqa
Note that, among the three $\hat v_i$, there are only two independent 
parameters. We thus study the constraints as a function of
\beqa
\sin\gamma&\equiv&\hat v_2\subset[0,1],\\
\sin\alpha&\equiv&\frac{\hat v_1}{\sqrt{\hat v_1^2+\hat v_3^2}}\subset[-1,1].\no
\eeqa
In terms of $\alpha$ and $\gamma$, we obtain
\beqa\label{eq:Zv2sa}
|z_1^K|&=&\Lambda_{12}^2
\left[\cos^2\gamma\sin^2\alpha+\sin^2\gamma\right],\\
|z_1^D|&=&\Lambda_{12}^2
\left[\cos^2\gamma\sin^2(\alpha-2\theta_c)+\sin^2\gamma\right],\no\\
{\cal I}m(z_1^K)&=&-\Lambda_{12}^2\sin\alpha\sin2\gamma,\no\\
{\cal I}m(z_1^D)&=&-\Lambda_{12}^2\sin(\alpha-2\theta_c)\sin2\gamma.\no
\eeqa

\begin{figure}[hbp]
\includegraphics[width=70mm]{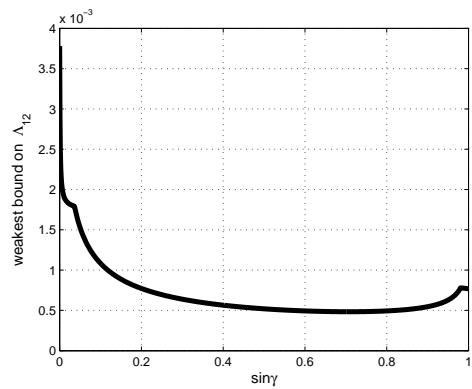}
\caption{The weakest $\Lambda_{12}$-bound as function of
  $\sin\gamma$.} 
\label{fig:sav2traj}
\end{figure}

As a first check of our results, note that when we take $\gamma=0$, we
reproduce Eq. (\ref{zonecpc}). (The identification of $\alpha$ with
$2\theta_d$ is correct only in the CPC case.)  The bound
(\ref{boucpc}) remains the weakest bound on the flavor degeneracy.
In the presence of a CPV phase in $V_d$, the bound becomes stronger.
The weakest $\Lambda_{12}$-bound as a function of $\sin\gamma$ is
presented in Fig. \ref{fig:sav2traj}.

At $0.03\lsim|\sin\gamma|\lsim0.98$, the constraints from the CPV
observables are dominant, and the combination of $z^{IK}_{\rm exp}$
and $z^{ID}_{\rm exp}$ is responsible for the unavoidable bound on
$\Lambda_{12}$. Defining 
\beq
r_{KD}^I\equiv z_{\rm exp}^{IK}/z_{\rm exp}^{ID},
\eeq
the weakest bound on $\Lambda_{12}$ corresponds to
\beq
\tan\alpha=\frac{r_{KD}^I\sin2\theta_c}
{1+r_{KD}^I\cos2\theta_c},
\eeq
and is given by
\beq
\Lambda_{12}^2\leq\frac{z_{\rm
    exp}^{ID}}{\sin2\theta_c\sin2\gamma}
{\sqrt{1+r_{KD}^{I2}+2r_{KD}^I\cos2\theta_c}}.
\eeq
Using Eq. (\ref{expim}), we find that the weakest bound occurs at
$\sin\alpha\approx0.014$
and it is given by
\beq\label{boucpv}
\Lambda_{12}\leq\frac{4.8\times10^{-4}}{\sqrt{\sin2\gamma}}
\left(\frac{\Lambda_{\rm NP}}{1\ {\rm TeV}}\right).
\eeq
Eq. (\ref{boucpv}) explains the $\sin\gamma$ dependence of the curve
in Fig. \ref{fig:sav2traj} in the relevant range.

Comparison with Eq. (\ref{boucpc}) reveals the power of the upper
bound on CPV in $D^0-\overline{D^0}$ mixing in constraining
the flavor structure of new physics. For maximal phases
($\sin2\gamma=1$), it implies degeneracy stronger by a factor of 8
compared to the bound from CPC observables. For $\Lambda_{\rm
  NP}\leq1$ TeV and large phases, the flavor-diagonal and
flavor-degeneracy factors should provide a suppression stronger than
${\cal O}(10^{-3})$. With loop suppression of order
$\lambda_{12}\sim\alpha_2$, the degeneracy should be stronger than
$0.02$.

\mysection{Supersymmetry}
An explicit example of the constraints on new physics parameters
obtained by combining measurements of $K^0-\overline{K^0}$ mixing and
of $D^0-\overline{D^0}$ mixing is provided by supersymmetry.  
Any supersymmetric model generates the operator (\ref{xqoperator}) via
box diagrams with intermediate gluinos and squark-doublets. The
various factors that enter $z_1^K$ and $z_1^D$ can be identified as
follows:
\beqa
\Lambda_{\rm NP}&=&\tilde m_Q\equiv(m_{\tilde Q_1}+m_{\tilde Q_2})/2,\no\\
\lambda_{12}^2&=&\frac{\alpha_s^2}{54}g(m_{\tilde g}^2/\tilde m_Q^2),\no\\
\delta_{12}&=&(m_{\tilde Q_2}-m_{\tilde Q_1})/(m_{\tilde
  Q_1}+m_{\tilde Q_2}), 
\eeqa
where $m_{\tilde Q_i}$ is the squark-doublet mass, $m_{\tilde g}$ is 
the gluino mass, and $g(m_{\tilde g}^2/\tilde m_Q^2)$ is a known
function (see {\it e.g.} \cite{Nir:2007xn}) with, for example
$g(1)=1$. Taking $\tilde m_Q\leq1$ TeV, and $m_{\tilde g}\approx
\tilde m_Q$ (which gives $\lambda_{12}\approx0.014$), leads to 
\beq\label{boususy}
\frac{m_{\tilde Q_2}-m_{\tilde Q_2}}{m_{\tilde Q_1}+m_{\tilde
    Q_2}}\leq \left\{ \begin{array}{cc}
    0.034 & {\rm maximal\ phases}\\
    0.27 & {\rm vanishing\ phases} \end{array} \right.
\eeq
We conclude that if squarks and gluinos are lighter than TeV, then the
first two squark doublets should be degenerate to, at least, order ten
percent. (Previous studies of the $\Delta m_K-\Delta m_D$ constraint
on the degeneracy between the squark doublets can be found in
Refs. \cite{Nir:2007xn,Feng:2007ke}.) 

\mysection{Warped extra dimension}
Another explicit example of the constraints on new physics parameters
is provided by the Randall-Sundrum (RS) type I framework~\cite{RS1}.
Allowing the SM fields to propagate in the bulk provides a solution to
the flavor puzzle~\cite{AS,original} and also partial protection
against flavor violation which is being induced, at tree level, by a
Kaluza-Klein (KK) gluon exchange process~\cite{aps}.  In models where
each SM doublet corresponds to a zero mode of a single 5d doublet,
operators of the form of (\ref{xqoperator}) arise.  Working in a two
generation framework and to leading order, we find~\cite{aps,RSEFT}
\beq
X_Q = {\rm diag}(f_{Q^1}^2,f_{Q^2}^2)\,, \ \ \Lambda_{\rm
  NP}^2\approx{3m_{\rm KK}^2\over 2\pi \alpha_s \xi }, 
\eeq
where $\xi\sim \ln(M_{\rm Pl}/{\rm TeV})\sim35$, and $f_{Q^i}$
corresponds to the value of the $i$th doublet field on the IR
brane. In this framework the hierarchy in the CKM mixing angles is
accounted for by a corresponding hierarchy in the $f_{Q^i}$s,
$f_{Q^1,Q^2}/f_{Q^3}\sim \lambda_C^3,\lambda_C^2$ with
$\lambda_C=0.23$, 
\beqa
\delta_{12}\approx 1\,, \ \ \lambda_{12}\approx {f_{Q^2}^2\over2}. 
\eeqa
We find the following constraints for the weakest bounds
\beq\label{fqtwo}
f_{Q^2}\leq \sqrt{m_{\rm KK}\over{\rm TeV}}\times
\left\{ \begin{array}{cc} 
    0.020 & {\rm maximal\ phases}\\
    0.056 & {\rm vanishing\ phases} \end{array} \right.
\eeq
There are two interesting limits to study. First, electroweak
precision measurements put a lower bound of $m_{\rm KK}\sim3$ TeV. At
the limit, our flavor constraints of Eq. (\ref{fqtwo}) imply
$f_{Q^2}\lsim 0.034$, which translates to $f_{Q^3}\lsim 0.67$. Second,
when the third generation doublet is fully composite, we have
$f_{Q^3}=1$. In this case Eq. (\ref{fqtwo}) gives a lower bound on the
KK scale: $m_{\rm KK}\gsim 6.7\,$ TeV. We conclude that for KK scale at
the LHC reach, CPV data  exclude the possibility of fully composite
third generation doublet (unless the CPV phase is accidentally small,
$\sin2\gamma\lsim0.2$). Note that in RS models flavor protection is
due to having $\lambda_{12}\ll 1$, while in the supersymmetry case it
is due to universality, $\delta_{12} \ll 1$.

\mysection{Conclusions}
The contributions of new physics to FCNC processes depend on four
features of the new physics: its scale $\Lambda_{\rm NP}$, a possible
flavor-blind suppression $\lambda_{12}$ (loop factors and kinematic
functions), flavor degeneracy $\delta_{12}$, and flavor alignment,
$\sin\alpha$. Low energy flavor measurements probe new physics only
indirectly. Usually, such indirect probes constrain only the product
of these four suppression factors, which is represented by the
coefficient of a non-renormalizable flavor-changing operator. In this
work we show, however, that some combinations of measurements go
beyond this simplest set of constraints and probe the new physics in
more detail.

In particular, for non-renormalizable operators that involve only
SU(2)-doublet quarks, the contributions to $K^0-\overline{K^0}$ mixing
and $D^0-\overline{D^0}$ mixing depend, to a good approximation, on
the same $\Lambda_{\rm NP}$-, $\lambda_{12}$- and
$\delta_{12}$-related factors but differ in their alignment factors in
a way that depends on the CKM matrix. Thus, the combination of these
measurements constrains, for TeV-scale new physics, the product of the
flavor-blind and flavor-degeneracy suppression factors [Eqs.
(\ref{boucpc}), (\ref{boucpv})]. Within supersymmetry, a strong
degeneracy between the first two squark-doublet generations is
required [Eq. (\ref{boususy})]. In models of warped extra dimensions,
a fully composite third generation quark doublet is generically
excluded, unless the KK scale is beyond the LHC reach
[Eq. (\ref{fqtwo})]. 

\mysections{Acknowledgments}
This research is supported by the United States-Israel Binational
Science Foundation (BSF), Jerusalem, Israel.  The work of YG is
supported by the NSF grant PHY-0757868. The work of YN is supported by
the Israel Science Foundation founded by the Israel Academy of
Sciences and Humanities, the German-Israeli foundation for
scientific research and development (GIF), and the Minerva Foundation.
The work of GP is supported by the Peter and Patricia Gruber Award.


\end{document}